\documentclass[a4paper,11pt]{article}
\pdfoutput=1 

\usepackage{amssymb}
\usepackage{jcappub} 

\usepackage[T1]{fontenc} 
\usepackage{dsfont}

\title{Averaging over cosmic structure: Cosmological backreaction and the gauge problem}

\author[a,b,1]{Dave B. H. Verweg,\note{Corresponding author.}}
\author[b]{Bernard J. T. Jones}
\author[b]{and Rien van de Weygaert}

\affiliation[a]{Bernoulli Institute for Mathematics, University of Groningen, The Netherlands}
\affiliation[b]{Kapteyn Astronomical Institute, University of Groningen, The Netherlands}

\emailAdd{dave@astro.verweg.net}

\abstract{We show that there are several implicit mathematical difficulties with numerous backreaction studies. These studies utilize perturbative or averaging methods. Regarding the former, we prove that the backreaction is a nested and multi-scaled effect, making the claim of Green \& Wald, asserting the backreaction to be negligible on large scales, to be untenable. For the averaging approaches, we present a mathematical treatment of spacetime slicing and its role in scalar averaging for cosmic structure. This reveals that averages are not invariant under foliation-gauge transformations. Consequently, different cosmic averages lead to different inferred results about the cosmic dynamics, resulting in artificial backreaction terms. Our analysis further demonstrates that the Gasperini's gauge-invariant averaging is mathematically ill-defined. To solve this gauge problem of averaging, we develop an averaging method that is invariant under scalar-preserving gauge-transformations.

\vspace{2mm}
}

\begin{document}
\maketitle
\flushbottom

\section{Introduction}
\label{sec:intro}
Over two decades ago, the Supernova Cosmology Project \cite{perlmutter1998discovery} and the High-Z Supernova Search Team \cite{riess1998observational} found that the expansion of the Universe was accelerating, suggesting the existence of a dominant 74\% contribution to the late-time energy distribution called \textit{dark energy} \cite{spergel2007three}. The requirement for its existence has been established by various studies \cite{efstathiou2002evidence, spergel2003wmap}, but its nature remains one of the biggest mysteries in cosmology up to date. Clarification of this problem is crucial as dark energy is relevant to almost all fields of cosmology—it is even understood as the overarching goal of precision cosmology \cite{wiltshire2008dark}. Dark energy not only dominates the expansion but also has substantial effects on the formation of highly nonlinear structure \cite{alimi2010imprints}, the cosmic microwave background (CMB) \cite{doran2001quintessence, caldwell2003early}, the dark matter distribution \cite{liberatorosenfeld2006darkenergy}, and the primordial structure growth \cite{ferreirajoyce1998scalingfield, bartelmann2005earlydarkenergy}.

The relation between dark energy and nonlinear structure dynamics seems peculiar at first glance. Structure formation dramatically impacts the composition of dark energy \cite{percival2005darkenergyandstructureformation}, and has thus been proposed as a sensitive probe for dark energy behavior \cite{liberatorosenfeld2006darkenergy}. This has been verified by \citet{alimi2010imprints}, who have shown that uniformly distributed and constant-in-time dark energy leaves an imprint on galactic dynamics on all scales and very dominantly in the nonlinear regime. Hence, large scale dynamics—in particular, the dark energy behavior—is affecting the local dynamics of nonlinear structures. It is captivating that dark energy started to dominate the Universe's energy density exactly around the time highly nonlinear structures began to form \cite{wetterich2002darkenergy, rasanen2004backreactionoflinear}, a phenomenon coined the \textit{coincide problem}. To many contributors, it therefore seemed natural to investigate the effect of nonlinear structures on the large-scale dynamics known as the \textit{backreaction effect}, see e.g. \cite{buchert2000averagedust, brandenberger2000backreaction, wetterich2003can, rasanen2004darkenergy, kolb2005effectofinhomogeneities, coley2005macroscopicgravity, li2008darkenergyeffectofaveraging}. We nonetheless know that local structure influences the metric, and, therefore, has an effect on galactic behavior \cite{brandenberger2000backreaction}. An explicit example of backreaction is that of gravitational waves \cite{brill1964method, isaacson1968gravitational, isaacson1968gravitational2}. The question is whether there are any dynamical effects of cosmic structure on the expansion of the Universe, and whether these backreaction effects are sizable or not. Some claim that the backreaction is negligible \cite{green2011framework}, while others claim it could be causing the cosmic acceleration \cite{buchert2000averagedust, brandenberger2000backreaction}.

In this paper, we look at the mathematical difficulties of methods that potentially allow us to capture dynamical effects of nonlinear structure on the cosmic expansion. Currently, the standard approach is either to take a perturbative or a spatial averaging approach. In Section \ref{section:current-methods}, we argue that perturbative models are ambiguous in capturing large-scale backreaction effects, but that working with cosmic averages seemed to have been proven useful. Section \ref{section:cosmic-averages-are-gauges}, however, shows that standard averaging approaches in backreaction studies do not pertain as they suffer from an implicit freedom of choice within fixing the foliation. Within this light, we discuss in Section \ref{section:averages-in-backreaction} the implications for backreaction studies, and show that averages lead to artificial backreaction quantification. Furthermore, we comment on an averaging procedure that has been understood as gauge invariant. To solve the gauge problem for averaging, we develop a perturbative averaging method in Section \ref{sec:perturbative-avg} that is invariant under scalar-preserving gauge-transformations. We conclude in Section \ref{sec:conclusion}.

In this paper, we solely consider classical gravity. Regarding notation, the Greek indices indicate the spacetime indices, whereas the Latin indices indicate the spatial ones.

\section{Perturbative and averaging methods for cosmological backreaction}\label{section:current-methods}
We discuss the current most prominent methods for quantifying dynamical structure effects on the cosmic expansion, which consist of either a perturbative or an averaging approach. We argue that the perturbative approach is unsuitable for quantifying such effects, but that averaging seems promising as numerical analyses portray convincing evidence. Furthermore, we explain why the most prevailing critique of the backreaction effect is misleading---that critique being that these effects should show up on Newtonian scales in our local Universe.

\subsection{Perturbative approaches}
Perturbative approaches are built upon realizing that inhomogeneities in the cosmic matter and energy distribution---such as clusters of galaxies---can be represented by deviations $\widetilde{g}_{\mu \nu}$, called \textit{perturbations}, of some smoothed-out metric $\overline{g}_{\mu \nu}$ which we employ as a background model:
\begin{equation}\label{eq:perturbative-decomposition}
g_{\mu \nu} = \overline{g}_{\mu \nu} + \widetilde{g}_{\mu \nu},
\end{equation}
see e.g. \cite{green2011framework, ishibashi2005acceleration}. Perturbative methods require a background model $\overline{g}_{\mu \nu}$ that contains all the large-scale properties for appropriate backreaction results \cite{cliftonsussman2019backreaction}. Backreaction effects are, therefore, representable by the effects of the inhomogeneities $\widetilde{g}_{\mu \nu}$ on the background $\overline{g}_{\mu \nu}$, that is, inferring the dependence relation:
\begin{equation}\label{eq:perturbative-backreaction}
\overline{g}_{\mu \nu}(\widetilde{g}_{\mu \nu}).
\end{equation}
This is a critical drawback of the perturbative methods as identifying the large-scale properties in the real model $g_{\mu \nu}$ with an appropriate background $\overline{g}_{\mu \nu}$ that solves the Einstein field equations is not yet known; see \citet{clarkson2011growthstructure}. And even if suitable backgrounds are identified, the choice of background hugely affects the physical results \cite{ellis1987fitting}, in particular, the backreaction \cite{cliftonsussman2019backreaction}. This is immediate from \eqref{eq:perturbative-backreaction} as the choice $\overline{g}_{\mu \nu}$ determines the backreaction representation $\overline{g}_{\mu \nu}(\widetilde{g}_{\mu \nu})$.

The convenience of describing backreaction effects by the influence of the perturbations $\widetilde{g}_{\mu \nu}$ on the behavior of the background $\overline{g}_{\mu \nu}$ has served as a cornerstone for perturbative methods analyzing these effects on large-scale properties, such as the expansion \cite{green2011framework}. In the face of the above argument, the averaging scale for smoothing the perturbed metric $g_{\mu \nu}$ to attain $\overline{g}_{\mu \nu}$ must be such that the large-scale effects are still present in the background \cite{zalaletdinov1992averaging, zalaletdinov1997averaging}. Otherwise, we would not be able to identify this background $\overline{g}_{\mu \nu}$ with the true large-scale properties of $g_{\mu \nu}$, that is to say \eqref{eq:perturbative-decomposition} does not capture the physical backreaction. As a consequence, these local effects on a large scale must be included in the background itself. Specifically, the averaged energy-momentum tensor $\overline{T}_{\mu \nu} \equiv \overline{T}_{\mu \nu} (g_{\mu \nu})$ captures the large-scale properties and their underlying effects, such as the backreaction. Intuitively, one can understand this by noting that backreaction effects cannot solely be identified as the difference $t^{(0)}_{\mu \nu}$ in
\begin{equation}\label{eq:perturbative-br-difference}
    G_{\mu \nu} ( \overline{g}_{\mu \nu} ) = 8\pi G \overline{T}_{\mu \nu} + t_{\mu \nu}^{(0)},
\end{equation}
where $t_{\mu \nu}^{(0)}$ is thought to be the energy-stress tensor of the backreaction \cite{green2016nobackreaction, bolejko2017current-status-inhomogeneous-cosmology}. However, it is imperative to note that all backreaction effects that impact large-scale properties, such as the cosmic expansion, must be encompassed within $\overline{T}_{\mu \nu}$. As a result, these effects are not accounted for in $t_{\mu \nu}^{(0)}$, as precisely indicated in the findings of \citet{green2011framework}. Notably, $t_{\mu \nu}^{(0)}$ solely captures the small-scale backreaction effects, which are solely gravitational waves. The fundamental reason behind this argument lies in the multi-scaled nature of backreaction, wherein it manifests itself across various scales of clustering and inhomogeneity.

To see this, consider a smooth family $g_{\mu \nu}^{(\lambda)}$ of metrics parametrized by a perturbation scale $\lambda > 0$, where $g_{\mu \nu}^{(0)} = \overline{g}_{\mu \nu}$, following standard practice in perturbative studies as has been popularized by \citet{green2011framework}. It has been suggested that the backreaction is likely to manifest over a range of $\lambda$ due to its multi-scaled nature \citep{buchert_ellis2015, ostrowski2017ongreenwaldformalism}. This is in contrast to the assumption made in previous studies \citep{green2011framework, green2012, green2013, green2014FLRW, green2016nobackreaction}, where backreaction effects were exclusively considered at $\lambda = 0$ in the form of \eqref{eq:perturbative-br-difference}. From a mathematical perspective, this can be proven based on the non-commutative nature of averaging and the calculation of the field equations \citep{ellis1984relativistic, ellis1987fitting, ellis2005universe}. Suppose that at some perturbation-scale $\lambda > 0$, the Einstein field equations hold:
\begin{equation}
G_{\mu \nu} \big( g_{\mu \nu}^{(\lambda)} \big) = \kappa T_{\mu \nu}^{(\lambda)}.
\end{equation}
We introduce an averaging $\langle \cdot \rangle_\varepsilon$, which smoothes out the cosmology represented by $g_{\mu \nu}^{(\lambda)}$ to an infinitely small order $\varepsilon > 0$. The non-commutation of averaging implies the existence of a non-zero tensor field $t_{\mu \nu}^{(\lambda - \varepsilon)}$ such that:
\begin{equation}
G_{\mu \nu} \bigg( \big\langle g_{\mu \nu}^{(\lambda)} \big\rangle_\varepsilon \bigg) = \bigg\langle G_{\mu \nu} \big( g_{\mu \nu}^{(\lambda)} \big) \bigg\rangle_\varepsilon + t_{\mu \nu}^{(\lambda - \varepsilon)}.
\end{equation}
Given that the averaging was selected such that $\langle g_{\mu \nu}^{(\lambda)} \rangle_\varepsilon = g_{\mu \nu}^{(\lambda - \varepsilon)}$, we explicitly obtain a backreaction term at the scale $\lambda - \varepsilon < \lambda$:
\begin{equation}\label{eq:pert-br-any-scale}
G_{\mu \nu} \big( g_{\mu \nu}^{(\lambda - \varepsilon)} \big) = \kappa T_{\mu \nu}^{(\lambda - \varepsilon)} + t_{\mu \nu}^{(\lambda - \varepsilon)}.
\end{equation}
Here, we have utilized the relationship $\langle G_{\mu \nu}(g_{\mu \nu}^{(\lambda)}) \rangle = 8 \pi G T_{\mu \nu}^{(\lambda - \varepsilon)}$, assuming that the averaging operation is well-defined \citep{ostrowski2017ongreenwaldformalism}. 

This proves that when smoothing out any degree of structure in a cosmology, an energy-stress backreaction term arises in the field equations, and it aligns with the notion that backreaction manifests itself across various scales of structure and its dynamics. It is important to note that this result disproves the validity of the assumption made by \citet{green2011framework} that one can construct a family $g_{\mu \nu}^{(\lambda)}$ such that only at $\lambda = 0$ there is a backreaction term $t_{\mu \nu}^{(0)}$.

An explicit example of multi-scale backreaction is presented by \citet{korzynski2015nonlineareffectsofgr}, wherein it is noted that it is the collective impact of inhomogeneity effects across all scales that leads to substantial deviations of the perturbed metric $g_{\mu \nu}$ from the smoothed-out background $\overline{g}_{\mu \nu}$. Therefore, equation \eqref{eq:pert-br-any-scale} holds for any non-zero $\varepsilon < \lambda$.

In the hypothetical case, however, where backreaction is presumed to be nonexistent at all smoothing scales $\lambda$ except in the limit $\lambda \to 0$, the mathematical formalism introduced in \cite{green2011framework} remains applicable. As these authors conclude, such small-scale perturbations solely influence the appropriate background through gravitational waves. Nonetheless, this perspective does not provide a means to quantify the multi-scaled dynamical effects of structure on large-scale properties, such as expansion.

\subsection{Averaging approaches}
Averaging approaches cover and divide an inhomogeneous cosmological model into compact regions and quantify local dynamics in each region by averaging over matter fields, such as the energy density, describing the content of the Universe; see e.g. \cite{zalaletdinov1992averaging, zalaletdinov1993towards, zalaletdinov1997averaging, buchert2000averagedust, buchert2001averageperfectfluid, buchert2020average}. For reviews, see \cite{buchert2008dark, wiltshire2008dark, wiltshire2009gravenergy, ellis2009dark, buchert2012backreaction}. To the best of our knowledge, averages in general relativity and cosmology have been introduced by \citet{noonan1984gravitational, zotov1992averaging} and \citet{futamase1988, futamase1989}. Note that averaging over the perturbed metric means smoothing the (local) cosmic inhomogeneities, which represent the complex dynamics and structuring of cosmic matter.

\subsubsection{The success of averaging in backreaction studies}
Numerous contributions using the averaging scheme provide numerical results that show the backreaction to be significantly influencing cosmological parameters \cite{kolb2005effectofinhomogeneities, kolb2006acceleration, enqvist2007ltb, marozziuzan2012anisotropy, adamek2015affectdynamics}. Prominent examples are:
\begin{itemize}
    \item[(a)] \citet{buchert2000cosmparamaters} who show that the cosmological density parameters, such as $\Omega_\Lambda$, in a region of 100 Mpc possibly can deviate more than 100\% under $3\sigma$-fluctuations in the initial cold dark matter density;
    \item[(b)] \citet{rasanen2004darkenergy} who shows the backreaction possibly can result in an equation of state $0 \leq \omega \leq -4/3$ without the need for dark energy;
    \item[(c)] \citet{adamek2018backreactionsimulations} who show that in highly perturbed regions, the backreaction contributions to the expansion rate can possibly be of order unity with comoving gauge time slicing.
\end{itemize}
To comprehend the meaning of these results, it must be noted that even if backreaction effects are small for sufficiently large regions, they can result in large deviations from the standard homogeneous values of the cosmological parameters on scales of 100 Mpc and smaller \cite{buchert2000cosmparamaters}. Since even small deviations in the cosmic matter distribution can lead to significant changes to the macroscopic description of a physical system \cite{ellis2005universe}, the above results imply a realistic possibility of the backreaction effect explaining the cosmic acceleration.

The significance of the above results on the backreaction effect has been criticized by either perturbative arguments \cite{green2011framework, green2014FLRW, green2016nobackreaction} or by Newtonian arguments. The former has been commented on above. The latter emerges from the idea that the local Universe is accurately described by Newtonian mechanics, since all velocities are small and all scales are small compared to the Hubble length $cH$ \cite{peebles1980large}. The backreaction appears to be non-existent \cite{kaiser2017there, buchert2018backreaction}, or at least negligible \cite{ishibashi2005acceleration, green2014FLRW}, in Newtonian theory. If backreaction effects are sizable, they must appear on Newtonian scales in the nonlinear regime, and thus especially in the local Universe. This has led to the conclusion that there is no backreaction on global scales---even not in the relativistic setting, cf. \citet{kaiser2017there}. Up to date, this appears to have confirmed the widely shared view on the backreaction within the scientific community: its effects are negligible on large-scale cosmic dynamics.

In recent years, it has become apparent that Newtonian dynamics does not adequately describe the local Universe in all relativistic scenarios, despite the scales and velocities involved being small and previously thought to be Newtonian. \citet{korzynski2015nonlineareffectsofgr}, for example, proves explicitly and rigorously that there are nonlinear effects caused by the way cosmic structure is nested and hierarchically distributed, and that these cannot be included within Newtonian gravity or first-order relativistic perturbation theory. Furthermore, it is shown that these effects can be very significant as the accumulation of nonlinear effects in general relativity can amplify the backreaction---if only enough different orders of scales are considered. The backreaction can consequently have sizable impacts on physical situations we have relied on to be almost perfectly Newtonian, ranging from structure formation and $N$-body simulations to determining the Hubble constant \cite{racz2017withoutDarkEnergy, wiltshire2008dark}.

The most prominent formalism up to now to describe backreaction effects, and in particular these multi-scale effects due to the clustering and nesting of structure, on large-scale cosmic dynamics seems to be that of averaging. In the following sections, we provide the mathematical background of cosmic averaging with respect to the spacetime slicing, for which we identify a fundamental problem. We first recall the meaning of spatial averaging in a novel way to set the appropriate context.

\subsubsection{What is averaging in cosmological context?}
We recapitulate the meaning of spatial averaging within cosmology. On the largest scales, the matter in the Universe shapes a knotty and tangled structure known as the \textit{cosmic web} \cite{bond1996nature}. The web has clusters of galaxies as nodes that are connected by filaments and sheets, forming the biggest nonlinear structures in the Universe. It is mainly dominated by the near-empty regions called \textit{voids} \cite{aragon2010multiscale}, and is characterized by the cosmic matter and energy distribution consisting of different fields that interact with one another; such as the baryonic, neutrino, and dark matter fields. For scales of up to 200 h${}^{-1}$Mpc, these \textit{matter fields} are exceedingly complex and nonlinear as they account for the highly perturbed density fluctuations. These complexities within matter fields can geometrically be thought of as a landscape of peaks, hills, and mountains built upon one another at multi-scale levels within the cosmic web. To analyze cosmic dynamics locally, it is crucial to incorporate the nestedness of structures, as it is a distinctive feature of the nonlinear characteristics that we seek. The problem is that, even at second-order, numerical and analytic computations for such high-resolution matter fields are extremely complicated due to the highly nonlinear terms, which are needed to maintain the exactness of the computations while dealing with the large degrees of freedom of such a high-resolution system.

Due to the complexity of the computations, it is natural to try to derive an auxiliary cosmological model with representative matter fields that approximate the cosmic matter distribution sufficiently well in order to provide relevant results without maintaining the same model resolution. There is thus a trade-off between the simplicity of analytic and computational results and the resolution of the model---the latter being a measure of the detail within the representation of the cosmic structure, see \citet{ellis1987fitting}. Working with a homogeneous model, for example, where most nonlinear structure has been smoothed out, results in a lower quality of results since the low-resolution model contains much less information than an inhomogeneous cosmology refined in accordance with our observable Universe. In other words, we want to reduce some of the complexity that comes with the highly nonlinear dynamics but preserve the most significant features. To achieve that goal of tuning the resolution of the cosmic web model, we can specify a length scale such that the auxiliary cosmology, and thus the inhomogeneities in the cosmic web, are represented realistically down to that scale \cite{ellis1987fitting}. In practice, we introduce an averaging method to average out the Universe over the desired distance scale.

For these reasons, it is appropriate to employ an averaging scheme on top of a $\Lambda$CDM background. In general, averaging a matter field $\phi$ over a region $\mathcal{D}$ means assigning a scalar value $\langle \phi \rangle_{\mathcal{D}}$ based on the content of the matter field in that region measured by a given volume form. If we average over every possible region bounded by an averaging scale, we can construct a field $\bar \phi$ that has relatively less complex cosmic structure than $\phi$ itself, and in that sense $\bar{\phi}$ is understood to be smoother than $\phi$. This averaged-out field $\bar \phi$ represents $\phi$ down to the averaging scale. Combining all the averaged matter fields, we can construct an averaged-out cosmological model via the Einstein field equations, which forms a landscape of smooth hills and mountains representing the clustering of structure in a smoothed-out way. This smoothed-out model is much easier to do analysis and computations on, and is understood to still maintain the large-scale properties of the inhomogeneous complex model representing our Universe characterized by the averaging scale \cite{ellis2005universe}.

The question of constructing an appropriate averaging method arises naturally when dealing with cosmic dynamics. However, in the case of a curved space, averaging is not geometrically defined as there is not a unique way to average nonlinear fields, cf. \cite{buchert_Ehlers1995averaging, gasperini2009gauge, brannlund2010averaging}. Below, we show that averaging turns out to implicitly induce gauges. This presents a challenge in accurately averaging the local dynamics of the cosmic web.

\section{Cosmic scalar averages are gauges}\label{section:cosmic-averages-are-gauges}
As discussed above, the literature suggests that averaging holds promise as a method for describing the cosmic dynamical effects on the expansion. Here, we provide a mathematical treatment of spacetime slicing and elucidate its role in cosmic averaging. This treatment brings to light a fundamental problem of these averaging procedures: selecting an averaging method necessitates specifying a spacetime-slicing, for which averaging is not invariant under its transformations. Consequently, the choice of averaging is not inherent to the cosmological model; that is, they are gauges. This limitation renders standard averaging non-representative for cosmic structure analyses.

    \subsection{Foliations are gauges}\label{section:foliations-are-gauges}
    We elucidate the precise manner in which these foliations serve as gauges, and we demonstrate that transitioning from one foliation to another typically impacts the representation of the cosmological model.

    \subsubsection{Foliation as representation of cosmology equivalence class}
    We demonstrate that choosing a foliation uniquely determines the metric tensor, which is a particular representation of the class of equivalent cosmological models. This gives rise to a rigorous definition of a gauge within cosmology.

\begin{figure}[h!]
        \centering
        \includegraphics[width=15cm]{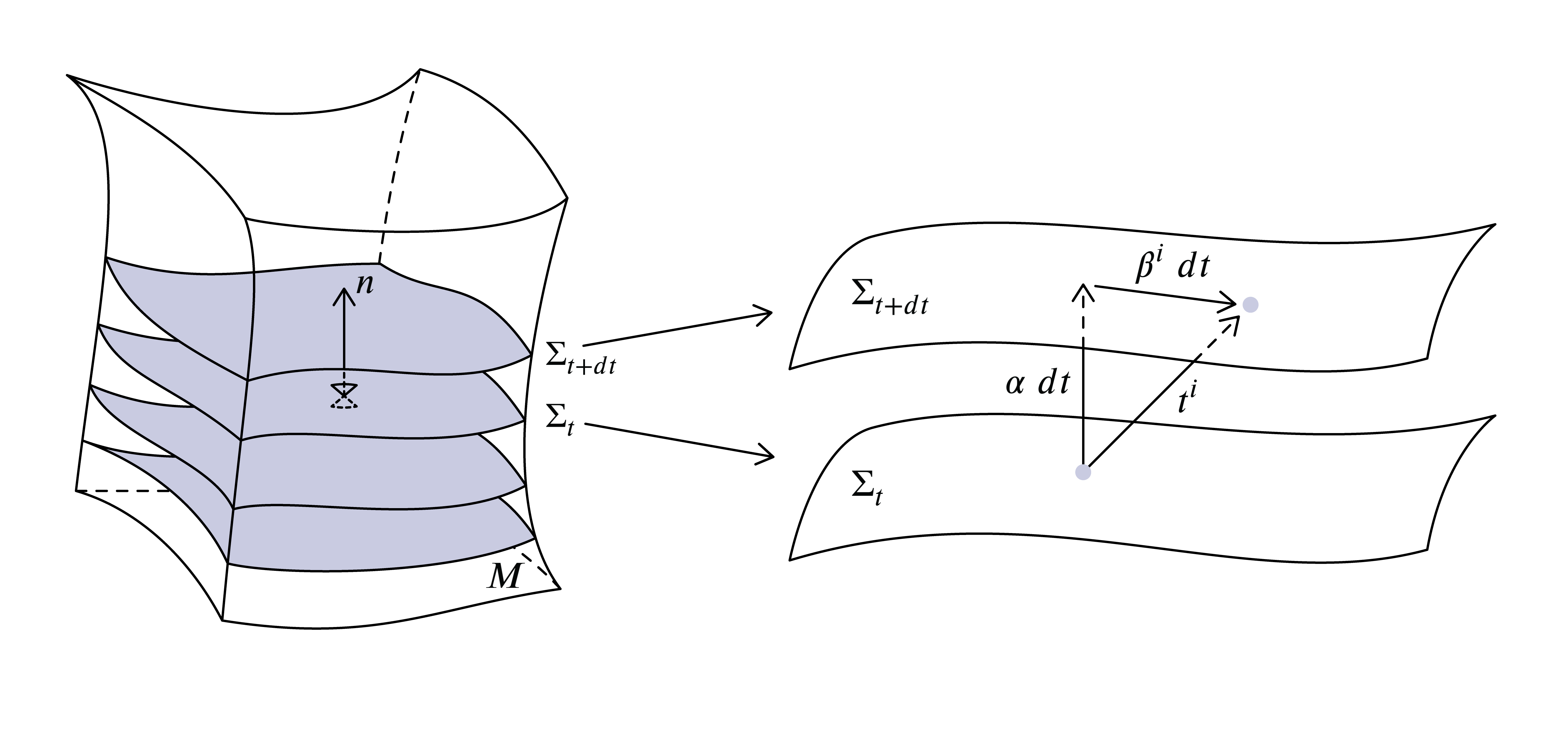}
        \caption{Sketch of foliation in ADM formalism.}
        \label{fig:foliation}
    \end{figure}

    First, we prove that selecting a foliation determines a unique representation of the metric tensor from its equivalence class under diffeomorphisms. In cosmology, it is usual practice to assume that the Universe can be modeled by a globally hyperbolic spacetime, implying that it admits a diffeomorphism $T: \mathbb{R} \times \Sigma \to M$ such that $M = \bigcup_{t \in \mathbb{R}} \Sigma_t$ with $\Sigma = \{\Sigma_t\}$ a disjoint set and $\Sigma_t := T_t(\Sigma)$ is spacelike for every $t \in \mathbb{R}$ where we write
    \begin{align}
        T_t: \Sigma \longrightarrow M, \qquad \qquad T_t (x) := F(t,x),  \\
        T_x: \mathbb{R} \longrightarrow M, \qquad \qquad T_x (t) := F(t,x).
    \end{align}
    Then we can define the lapse function $\alpha$ and shift vector $\beta^i$ by
    \begin{equation}
        \frac{dT_t}{dt} =: \alpha \boldsymbol{n} + \boldsymbol{\beta},
    \end{equation}
    where $\boldsymbol{n}$ is the unit normal to the foliation $\Sigma$ \cite{landsman2021foundations}; see Figure \ref{fig:foliation}. For a standard treatment on the ADM formalism for numerical relativity, see for example \citet{corichi_intro-to-ADM}. In general coordinates $x^\mu$ adapted to the foliation, this completely determines the 4-metric $g_{\mu \nu}$ on $M$ as the ADM formalism tells us that
    \begin{equation}
        g_{\mu \nu} dx^\mu dx^\nu = - \alpha^2 dt^2 + h_{ij} (dx^i + \beta^i dt)(dx^j + \beta^j dt),
    \end{equation}
    with $h_{ij}(t, \cdot)$ is the spatial metric on slice $\Sigma_t$.

    Second, we address the correspondence between the metric tensor and the foliation, which allows us to clarify the implications of choosing a foliation in terms of how it specifies a representation of equivalent cosmologies. The gauge invariance in cosmology arises from the fact that a cosmological model $(M, g, \{F_i\})$ with matter fields $\{F_i\}$ is physically equivalent to a model transformed by an orientation-preserving diffeomorphism, see \citet{hawkingellis1973}. For the collection of all such equivalent models, we write
    \begin{equation}\label{eq:equiv_class}
        \big[ (M, g, \{F_i\})  \big] := \bigg\{ (M, g', \{ F'_i \}) \; \big| \;   \varphi_* g = g', \; \varphi_* F_i = F'_i \; \text{for all $i$} \bigg\},
    \end{equation}
    where $\varphi \in \text{Diff}(M)$. The cosmological model is thus described by $[(M,g, \{F_i \})]$, and for calculations we select a representation from this equivalence class. Such representation is formally called a \textit{gauge}. As shown above, specifying a foliation $\Sigma$ is equivalent to selecting a representation $g \in [g]$ on spacetime $M$. Therefore, there is a one-to-one correspondence between each equivalent foliation in $[\Sigma]$ and the metric tensors in $[g]$. Since a foliation $\Sigma$ is always such that $M = \bigcup_{t \in \mathbb{R}} \Sigma_t$, a representation $(M, g, \{F_i\})$ from the equivalence class \eqref{eq:equiv_class} of cosmologies is in unique correspondence with the foliation $(\Sigma, \{F_i\})$. 
    
    Consequently, a gauge transformation $\varphi \in \text{Diff}(M)$ that induces $g \mapsto g'$ can be represented by a foliation transformation
    \begin{equation}
            \big(\Sigma, \; \{F_i\}  \big) \longmapsto \big(\Sigma', \; \{F'_i\} \big),
    \end{equation}
     In short, we write $\Sigma \mapsto \Sigma'$.  Here, the identification of the correctly determined matter fields $\{ F'_i \}$ with the transformed foliation $\Sigma'$ in the representation $(\Sigma', \; \{F'_i\} )$ from the equivalence class $\big[ (\Sigma, \; \{F_i\}) \big]$ is of importance for an argument below.

    \subsubsection{Foliation-gauge transform is a change in representation}
    In order to highlight the criticality of the foliation choice as a gauge, we present a proof showcasing that transitioning from one equivalent foliation to another typically brings about a fundamental modification in the representation of the cosmological model, rather than a mere relabeling of points. It is important to note the distinction between the cosmological model itself and its representation, as a foliation change does not impact the model itself \cite{hawkingellis1973}. 
    
    To see this, note that locally a diffeomorphism $\varphi \in \text{Diff}(M)$ on a chart $U \subset M$ with coordinate map $\psi: U \to \mathbb{R}^4$ can be written as
    \begin{equation}
        x := \psi(p) \longmapsto \psi\big( \varphi(p) \big) =: y,
    \end{equation}
    for any $p \in U$. Let $v \in TM$ a tangent vector field over $U \subset M$, which we can decompose as $v=v^\mu \partial_{x^\mu}$. Under the foliation change $\varphi \in \text{Diff}(M)$, the inner product of $v$ gets transformed as
    \begin{equation}\label{eq:innerproduct-transformation}
        g_{\mu \nu}^{(x)} v^\mu v^\nu \longmapsto \bigg( \frac{\partial x^\alpha}{\partial y^\mu} \frac{\partial x^\beta}{\partial y^\nu}  g_{\alpha \beta}^{(x)} \bigg) \bigg( \frac{\partial y^\mu}{ \partial x^\alpha } \frac{\partial y^\nu}{ \partial x^\beta} v^\alpha v^\beta \bigg) 
        = \varphi_* \big( g_{\mu \nu}^{(x)} v^\mu v^\nu \big),   
    \end{equation}
    where the metric dependence on the coordinates is denoted explicitly by $g_{\mu \nu}^{(x)}$. If $\varphi$ induces a Lorentz transformation, then 
    \begin{equation}
        \varphi_* \big( g_{\mu \nu}^{(x)} v^\mu v^\nu \big)
        = g_{\mu \nu}^{(y)}  \bigg( \frac{\partial y^\mu}{ \partial x^\alpha } \frac{\partial y^\nu}{ \partial x^\beta} v^\alpha v^\beta \bigg) 
        = g_{\alpha \beta}^{(x)} v^\alpha v^\beta.
    \end{equation}   
    That is the well-known result that \eqref{eq:innerproduct-transformation} is invariant under Lorentz transformations. However, under a general gauge transformation $\varphi \in \text{Diff}(M)$, we have 
    \begin{equation}\label{eq:gt-line-element}
        \varphi_* \big( g_{\mu \nu}^{(x)} v^\mu v^\nu \big)
        = g^{(x)}_{\mu \nu} v^\mu v^\nu + \sum_\alpha \sum_{\beta \neq \alpha} g^{(x)}_{\alpha \beta} \frac{\partial x^\alpha}{\partial y^\mu} \frac{\partial x^\beta}{\partial y^\nu} 
        \bigg( \sum_{  (\mu, \nu) \neq (\alpha, \beta) } \frac{\partial y^\mu}{\partial x^\alpha } \frac{\partial y^\nu}{ \partial x^\beta} v^\mu v^\nu  \bigg),
    \end{equation}   
    where this last term is generally nonzero, and therefore \eqref{eq:gt-line-element} does not equate to $g^{(x)}_{\mu \nu} v^\mu v^\nu$. Note that this result is in no way contradicting the principle of general covariance as \eqref{eq:innerproduct-transformation} maps to 
    \begin{equation}
        \varphi_* \big( g_{\mu \nu}^{(x)} v^\mu v^\nu \big) = g_{\mu \nu}^{(y)} \tilde{v}^\mu \tilde{v}^\nu,
    \end{equation}
    with $v = \tilde{v}^\mu \partial_{y^\mu}$. 
    
    It is crucial to remember that the spacetime interval serves as the fundamental property of general relativity, quantifying the metric tensor and playing a central role in the theory. The metric tensor fully specifies the geometry of spacetime, being the only quantity from which the laws of relativity can be derived, see \citet{wald1984GR}. As a result, a foliation transformation $\Sigma \mapsto \Sigma'$ induces a fundamental change in the representation of a cosmology. It is a choice of `rest-frame' and simultaneity over all spacetime points. This observation aligns with the relativistic understanding that different selections of rest-frames lead to varying perceptions of simultaneity, which adheres to the equivalence principle. Thus, foliation transformations do not alter the cosmological model itself but do alter its representation. Note that when undertaking practical analyses and computations, averaging necessitates the selection of a specific representation.

    In summary, the preceding discussion demonstrates that fixing a foliation is equivalent to selecting a metric representation from its equivalence class under diffeomorphisms, constituting a gauge choice. This choice is intrinsically related to the representation of simultaneity in relativistic models, as evidenced by its influence on the representation of the spacetime interval. The foliation choice assumes critical importance when averaging, raising questions about how this gauge choice impacts the conventional practice of averaging matter fields. This is particularly true in the context of cosmic backreaction.

    \subsection{The averaging dependence on foliation choice}\label{section:nonequivalent_averages}
    The treatment above revealed that specifying a foliation is a gauge choice with respect to the cosmology. We show that standard scalar averaging is foliation-dependent, and thus gauge-dependent.

    \subsubsection{Local-coordinate argument}
    Here, we follow \citet{gasperini2009gauge} with focus on the role of the foliation.
    
    Fix a domain $\mathcal{D} \subset M$. The above showed that any gauge transformation $\varphi \in \text{Diff}(M)$ is identified with a foliation change $\Sigma \mapsto \Sigma'$ denoted by $\varphi$. We are interested in the effect of $\varphi$ on the standard cosmic averaging procedure,
    \begin{equation}\label{eq:scalar-avg}
        \frac{\int_{\mathcal{D}} S(x) \sqrt{-g (x)} \; d^4 x}{\int_{\mathcal{D}}  \sqrt{-g (x)} \; d^4 x}.
    \end{equation}
    The integral over $S$ transforms as
    \begin{equation}\label{eq:avg-gauge-transformed}
        \int_{\mathcal{D}} S(x) \sqrt{-g (x)} \; d^4 x
        \longmapsto
        \int_{\mathcal{D}} S'(x) \sqrt{-g' (x)} \; d^4 x,
    \end{equation}
    where the apostrophe denotes the transformed fields. Note that a scalar field $S$ and the square root of the metric determinant transform as
    \begin{equation}\label{eq:metric-determinant-transform}
        S(x) \longmapsto S' (x) = S \big( \varphi^{-1}(x) \big); \qquad \sqrt{-g(x)} \longmapsto \bigg| \frac{\partial \varphi}{\partial x} \bigg|^{-1} \sqrt{-g \big( \varphi^{-1}(x)} \big).
    \end{equation}
    These expressions can be used to retrieve the effect of the foliation change in \eqref{eq:avg-gauge-transformed}, by first making a global diffeomorphism transformation, which we denote locally by $x \mapsto \bar{x} := \varphi^{-1}(x)$. That is,
    \begin{align}
        \int_{\mathcal{D}} S(x) \sqrt{-g (x)} \; d^4 x
        \quad \longmapsto \quad 
        & \int_{\varphi^{-1}(\mathcal{D}) } S'\big( \varphi(\bar{x}) \big) \sqrt{-g'\big(\varphi(\bar{x}) \big)} \; \bigg| \frac{\partial \varphi}{\partial \bar{x}}\bigg| \; d^4 \bar{x} \\
        = & \int_{\varphi^{-1}(\mathcal{D})} S(\bar{x}) \sqrt{-g(\bar{x})} \; d^4 \bar{x}.
    \end{align}
    Evidently, the scalar averaging \eqref{eq:scalar-avg} is not invariant under an arbitrary foliation change $\Sigma \mapsto \Sigma'$ as the effect of the foliation can we understood as a deformation of the integration domain $\mathcal{D}$ to $\varphi^{-1}(\mathcal{D})$, while leaving everything else fixed.

    Since the foliation choice corresponds to how we establish an inertial frame of reference at each point in spacetime, the averaging process relies on the choice of simultaneity. The laws of physics remain unaltered after a foliation change in each reference frame, but the reference frames do get transformed themselves.

    \subsubsection{Coordinate-independent argument}\label{sec:coord-indep-argument}
    While the derivation in terms of local coordinates clarifies the effect of a gauge transformation on the integration domain with respect to the foliation, it fails to elucidate the nature of the transformation itself. Here we employ a coordinate-independent approach that unveils the foliation transformation, providing a clear understanding of where the gauge dependence originates within the averaging process.

    A gauge transformation can be written as $x^\mu \mapsto x^\mu + \xi^\mu$ for arbitrary scalar functions $\xi^\mu$, see \citet{bardeen1980gauge, mukhanov1992theory}. The foliation $\Sigma$ transforms to $\Sigma'$, which is fully determined by how the lapse $\alpha$, shift $\beta^i$ and the 3-metric $h_{ij}$ transform. The induced foliation transformation equations on scalar averages follow from the fact that any scalar field $S:M \to \mathbb{R}$ is transformed as
    \begin{equation}
        S \longmapsto S + \mathcal{L}_\xi S,
    \end{equation}
    where $\mathcal{L}_\xi$ is the Lie derivative along $\xi$. The integral over $S$ transforms as
    \begin{equation}\label{eq:coord-indep-transformed}
        \int_{\mathcal{D}} S \sqrt{-g} \; d^4x \longmapsto
        \int_{\mathcal{D}} \bigg( S + \mathcal{L}_\xi S \bigg)  \bigg( \sqrt{-g} + \mathcal{L}_\xi \sqrt{-g} \bigg) \; d^4 x.
    \end{equation}
    Since $\mathcal{L}_\xi \sqrt{-g} = - \frac{1}{2} \mathcal{L}_\xi g /  \sqrt{-g}$, we can retrieve a perturbed form of the transformed integral in \eqref{eq:coord-indep-transformed}, 
    \begin{equation}\label{eq:coord-indep-transformed2}
        \int_{\mathcal{D}} S \sqrt{-g} \; d^4x \longmapsto
        \int_{\mathcal{D}} S \sqrt{-g} \; d^4 x  +  \int_{\mathcal{D}} \bigg(  \sqrt{-g} \mathcal{L}_\xi S - \frac{1}{2} \big( S + \mathcal{L}_\xi S \big)\mathcal{L}_\xi g / \sqrt{-g}  \bigg) d^4 x.
    \end{equation}
    
    The above derivation provides insight into the averaging dependence on foliation-gauges. To derive this, note that the last integral term in \eqref{eq:coord-indep-transformed2} must vanish for the averaging procedure to be gauge-invariant. Since this must hold for any domain $\mathcal{D}$ we select, the integrand itself must be zero. So, we can deduce that if we were to select a representation foliation $\Sigma \in [\Sigma]$ such that the scalar averaging stays invariant under gauge transformation generated by $\xi^\mu$, then the determinant of the metric must satisfy
    \begin{equation}
        g = - \frac{1}{2} \bigg( 1 +  \frac{S}{\mathcal{L}_\xi S} \bigg),
    \end{equation}
    while preserving the volume integral.\footnote{We assume that $S$ is such that $\mathcal{L}_\xi S \neq 0$ in $\mathcal{D}$, else the averaging is already not gauge-invariant as the last integral-term in \eqref{eq:coord-indep-transformed2} is, in that case, generally non-zero.} If so, the last term in \eqref{eq:coord-indep-transformed2} vanishes. However, we see that this condition on $g$ depends on $\xi$ that generates the gauge transformation, making the general 4-averaging \textit{intrinsically} gauge-dependent.

\section{The impact of averaging on cosmic backreaction studies}\label{section:averages-in-backreaction}
Having treated how cosmic averaging depends on the spacetime slicing, we now specify our treatment to the backreaction studies. We show that the foliation-gauge fixing leads to artificial backreaction terms. Additionally, we comment a gauge-invariant averaging procedure that has been proposed in the literature in light of the above findings.

    \subsection{Spatial averages induce artificial backreaction terms}
    Regarding the aforementioned averaging dependence on foliation-gauges, we provide an example that highlights the intricacies and practical significance of this gauge dependence in constructing an averaging procedure. We first work out a simple cosmological setting for a foliation change affecting the prior chosen averaging procedure. Next, we compare the backreaction results that are found when utilizing these two averaging procedures.
    
    \subsubsection{The 3-averaging and 4-averaging are equivalent up to a foliation change}    
    Consider the standard scalar 3-averaging procedure,
    \begin{equation}\label{eq:avg-4}
        \langle S \rangle^{(3)}_\mathcal{D}(t) = \frac{\int_{ \mathcal{D}} S(x) \sqrt{h (x)} \; d^3 x}{\int_{p \in \mathcal{D}}  \sqrt{h (x)} \;W d^3 x},
    \end{equation}
    with $h$ the determinant of 3-metric $h_{ij}$ and $S:M \to \mathbb{R}$ a scalar field. By gauge-transforming\footnote{Indeed, $\alpha^{-1} \cdot \text{id}$ is a valid diffeomorphism as $\alpha>0$ ensures that the casual structure is preserved.} the 3-average by $\alpha^{-1} \cdot \text{id} \in \text{Diff}(M)$ with lapse function $\alpha$, we get
     \begin{equation}\label{eq:avg-3}
        \langle S \rangle^{(3)}_\mathcal{D} (t) \longmapsto \langle S \rangle^{(4)}_\mathcal{D} (t) =  \frac{\int_{\mathcal{D}} S(x) \sqrt{-g (x)} \; d^3 x}{\int_{ \mathcal{D}}  \sqrt{-g (x)} \; d^3 x},
    \end{equation}
    as $\sqrt{-g} = \alpha \sqrt{h}$, and thereby retrieving the 4-averaging over spatial domain $D \subset \Sigma_t$. Although the 4-averaging and 3-averaging has been widely utilized in cosmological studies, their correspondence in light of gauges has not been pointed out: they are equivalent up to a foliation transformation. In a sense, we thus see that the spatial averaging with the 3-metric or the 4-metric is just a choice of foliation-gauge. It is the 3-averaging that has been popularized in backreaction studies by \citet{buchert2000averagedust}. It has been noted by \citet{gasperini2009gauge} that \eqref{eq:avg-4} and \eqref{eq:avg-3} do generally not give similar results for a scalar field $S$. Here we see it is an implicit gauge choice in the weighing of scalar field contributions within the averaging, because of its foliation dependence.

    \subsubsection{Example of local acceleration due to artificial backreaction term}
    We examine the significance of the foliation dependence in the context of the cosmological backreaction by exploiting an example: comparing \textit{inferred} backreaction results in terms of 4-averaging with those of the 3-averaging. 
    
    Consider a general inhomogeneous cosmology acting like an irrotational perfect fluid, which is globally hyperbolic. The Raychaudhuri equation describes the local spatial expansion $\theta$ with respect to evolution of time as measured by comoving observers $\xi$,
    \begin{equation}\label{eq:raychadhuri}
        \dot{\theta} = -\frac{\theta^2}{3} -2 \sigma + 2 \omega^2 - R_{ab} \xi^a \xi^b + {\dot{\xi}^a}_{;a},
    \end{equation}
    where $\sigma$ is the shear and $\omega$ the vorticity. Employing $\langle \cdot \rangle^{(i)}$ with $i=3,4$, one can deduce a positive term $\mathcal{Q}^{(i)}$ by averaging \eqref{eq:raychadhuri}, contributing to the averaged-out acceleration evolution $ \partial_t^2 \langle a \rangle^{(i)} / \langle a \rangle^{(i)}$ in region $D(t)$. For short, we suppress denoting the general expression $(i)$. As \citet{buchert2000averagedust} showed that, even in the simple case of a dust cosmology, this `backreaction' term $\mathcal{Q}$ shows up, 
    \begin{equation}
        \frac{ \partial_t^2 \langle a \rangle }{ \langle a \rangle} = - \frac{4}{3} \pi G \langle \rho \rangle + Q,
    \end{equation}
    In accordance with this formulation, \citet{kolb2006acceleration} deduce that for the averaged expansion in a region $D$ to be \textit{accelerating} if
    \begin{equation}\label{eq:acceleration-condition}
        \mathcal{Q} > 4 \pi G \langle \alpha^2 (\rho + 3p) \rangle,
    \end{equation}
    with $\rho$ the mass density, supposedly revealing the nature of the backreaction term $\mathcal{Q}$ in dark energy behaviour. Here we generalized this `acceleration condition' for a dust to perfect fluid cosmology, analogous to the work in \cite{buchert2001averageperfectfluid}.

    We work out an example for an irrotational perfect cosmological fluid. Consider a relatively small region $\mathcal{D}(t)$ within a void of, say, 5 Mpc in diameter, which allows us to validly describe its dynamics by the Newtonian limit as the gravitational potential energy is small, as well as the peculiar velocities \cite{peebles1993physicalcosmology}. We work out the Newtonian limit of the `acceleration condition'. To see this, note that we can write \eqref{eq:acceleration-condition} as 
    \begin{equation}
        \langle \alpha \theta \rangle^2 < 18 \pi G \langle \alpha^2 \rho \rangle - \frac{3}{2} \langle \alpha^2 R \rangle,
    \end{equation}
    with $\theta^2 = 24 \pi G \rho - 3 R/2$ and $R := h^{ij}R_{ij}$ the spatial curvature scalar, because by \cite{buchert2000averagedust, buchert2001averageperfectfluid} we have \begin{equation}
        \mathcal Q = 16 \pi G \langle \alpha^2 \rho \rangle - \langle  \alpha^2 R \rangle  - \frac{2}{3} \langle  \alpha \theta \rangle^2.
    \end{equation}
    In the Newtonian limit, the `acceleration condition' of the considered void region $\mathcal{D}(t)$ is
    \begin{equation}\label{eq:acc-condition-newton}
        \langle \alpha \sqrt{\rho} \rangle^2 < \frac{12}{18} \langle \alpha^2 \rho \rangle,
    \end{equation}
    as $R = 4 \pi G \rho$. By help of a toy model, we showcase that \eqref{eq:acc-condition-newton} need not hold at the same time for the 3-averaging and the 4-averaging.

    \subsubsection{Case study: artificial backreaction term in voids}
    We examine a simplified example that serves as a toy model for analyzing the `acceleration' condition prescribed by averaged backreaction terms. We continue in the Newtonian limit, as discussed above.
    
    Suppose that the region $\mathcal{D}(t)$ is located close to the boundary of the void near a filament, allowing us to set $\rho = x^1$ throughout $\mathcal{D}(t)$ for a particular time interval containing $t$, which we can do by orientating the void in the appropriate way. We are free to choose the foliation, and thus we can consider harmonic slicing in normal coordinates for which we can set $\alpha=x^1\sqrt{\det h_{ij}}$. In this case,
    \begin{equation}
        {\langle \alpha \sqrt{\rho} \rangle^{(3)}} < \sqrt{\frac{12}{18} \langle \alpha^2 \rho \rangle^{(3)} }, 
        \qquad
        {\langle \alpha \sqrt{\rho} \rangle^{(4)}} > \sqrt{\frac{12}{18} \langle \alpha^2 \rho \rangle^{(4)} }.
    \end{equation}
    That is, the considered void region is inferred to be accelerating when using averaging $\langle \cdot \rangle^{(3)}$, but decelerating when using $\langle \cdot \rangle^{(4)}$. In other words,
    \begin{equation}
        \frac{\partial_t^2 \langle a \rangle^{(3)}}{\langle a \rangle^{(3)}} > 0, \qquad \frac{\partial_t^2 \langle a \rangle^{(4)}}{\langle a \rangle^{(4)}} < 0,
    \end{equation}
    hold at the same time for the constructed region $\mathcal{D}$ at time $t$. 
    \begin{figure}[h!]
        \centering
        \includegraphics[width=15cm]{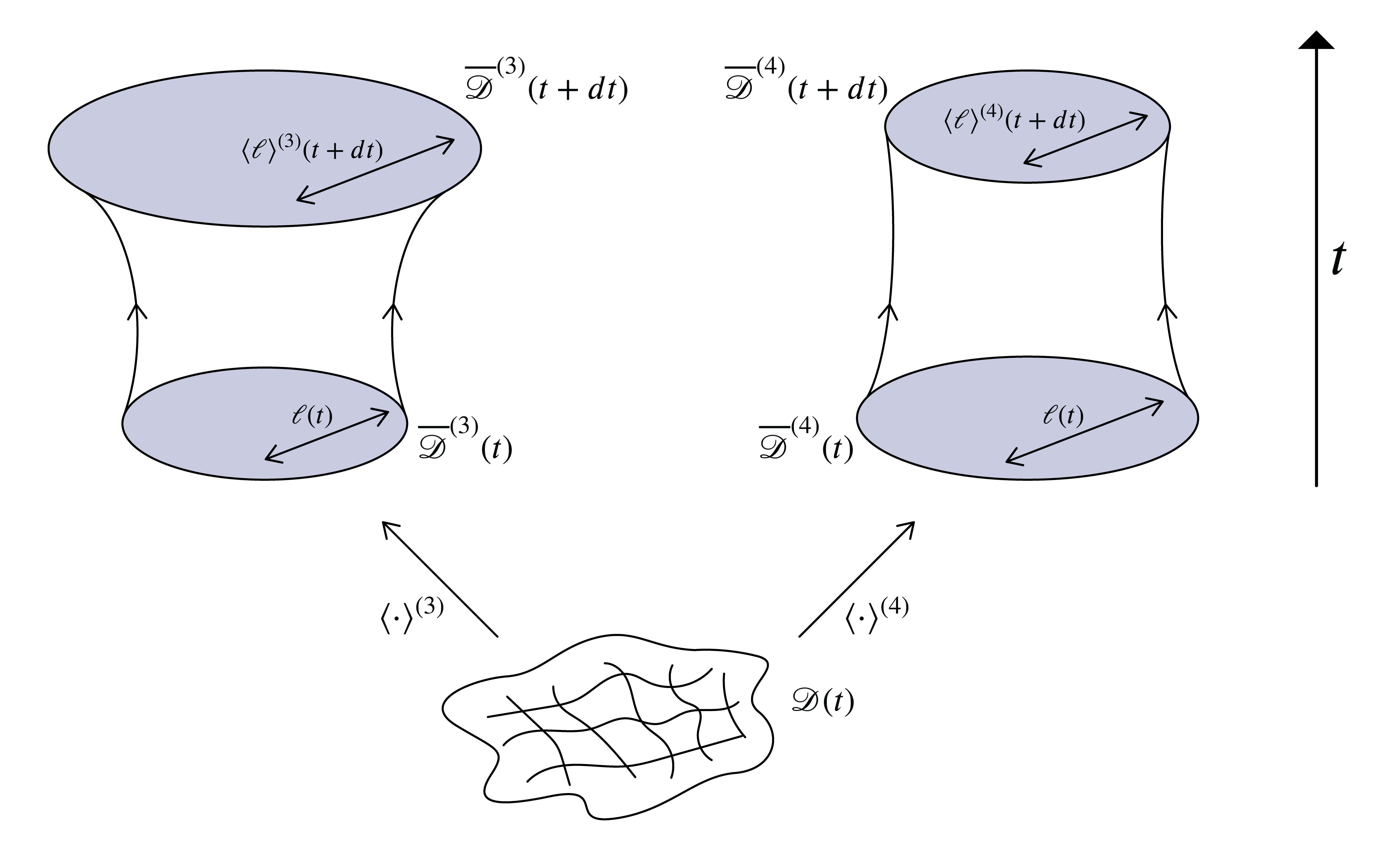}
        \caption{Sketch of the inferred expansion realities for the void region $\mathcal{D}(t)$.}
        \label{fig:acc-condition}
    \end{figure}
    
    Visually, this can be portrayed as in Figure \ref{fig:acc-condition}, where we define the averaged proper length by $\langle \ell \rangle (t+dt) := \ell(t) \partial_t  \langle a \rangle$ with $\ell(t)$ some initial proper length. The averaged length $\langle \ell \rangle (t+dt)$ is then used to infer the physical length $\ell(t+dt)$. The figure illustrates that when averaging the perturbed matter fields defined on $\mathcal{D}(t)$ with two different averaging procedures, this can give contradiction results: the averaged-out region $\overline{\mathcal{D}}^{(3)}$ might appear to be accelerating, while the other averaged-out region $\overline{\mathcal{D}}^{(4)}$ seems to be decelerating. 
    
    This example illustrates the intricate mathematical nuances inherent in the process of averaging---a fundamental underlying problem within the averaging procedure popularized by \citet{buchert2000averagedust, buchert2001averageperfectfluid}. Specifically, the aforementioned demonstrates the existence of scenarios in which the backreaction term $\mathcal{Q}$, derived via a particular averaging method, is artificial. The imperative notion dictates that these backreaction terms ought to be intrinsic to the cosmological model, uninfluenced by external choices. Contrarily, the averaged quantities explicitly hinge upon the chosen foliation, a consequence of the implicitly fixed representation of the cosmological construct during the formulation of the averaging procedure.

    In essence, this renders the averaged backreaction terms within the Buchert framework as discretionary variables, like coordinate systems that are easy to transform, yielding different outcomes. This gauge dependence of averaging potentially underlies the contradictory results of the ongoing debates surrounding backreaction and its plausible implications for the cosmic acceleration.

    \subsection{Gauge-invariant averaging: the Gasperini suggestion}
    The implication of our discussion, understanding gauges as foliation choices, leads to an elegant clarification of the intricacies that could arise from a particular specification of averaging. As an example, we consider an averaging method that has been introduced before; it has been suggested in order to solve the gauge dependence problem.
      
    \citet{gasperini2009gauge} note that scalar averaging $\langle \cdot \rangle^{(3)}$ is not invariant under the diffeomorphisms on $M$, which form the group $\text{Diff}(M)$. They proceed to construct an averaging procedure,
    \begin{equation}\label{eq:avg-procedure-gasperini}
        \langle S \rangle^\circ_\mathcal{D} (\tau) := \frac{ \int_{\mathcal{D}(\tau)}   \widehat{S}(\tau,x) \sqrt{\widehat{h} (\tau,x) } d^3 x  }{ \int_{\mathcal{D}(\tau)} \sqrt{\widehat{h}(\tau,x) } d^3 x  },
    \end{equation}
    where $\widehat S = S \circ \phi^{-1}$ and $\widehat h = h \circ \phi^{-1}$ with $\phi$ being the map $\phi(t, x) = \big(\phi^1(t,x), x \big) = (\tau, x)$ such that the foliation are now slices over $\phi(t, x) =\text{const}$.

    Considering the auxiliary map $\phi$ makes visible that $\langle \cdot \rangle^\circ$ weighs the contributions of $S$ in a particular way. The averaging procedure is constructed in such a way that it weighs the scalar field contributions over the slice of the transformed foliation $\phi(\Sigma)$, where we understand $\phi$ as a foliation change $\phi: \Sigma \to \phi(\Sigma)$. Notice the subtle point here: we are not weighing $S$ with respect to its implicitly fixed foliation $\Sigma$ but $\phi(\Sigma)$. This is problematic as $S$ can physically only be considered with respect to $\Sigma$, and not $\phi(\Sigma)$.
    
\begin{figure}[h!]
        \centering
        \includegraphics[width=15cm]{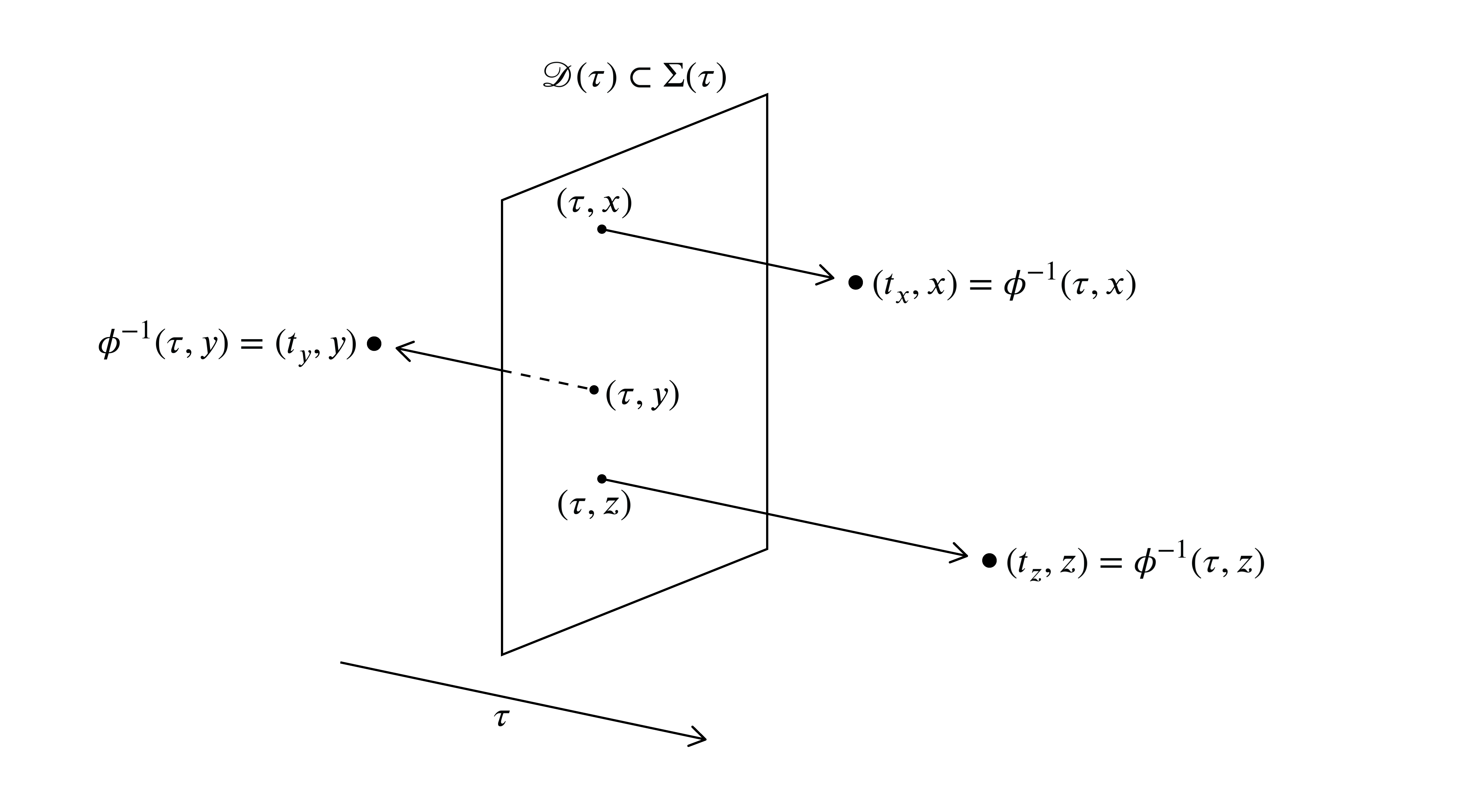}
        \caption{Sketch of foliation change $\phi^{-1}$ for $\mathcal{D}(\tau)$ in slice $\Sigma(\tau)$.}
        \label{fig:nonequivalent-averages-transformation}
    \end{figure}

    To see this, note that the foliation has been specified to be the family $\big\{ \Sigma( \phi^1(t,x) ) \big\}$, where one slice is of the form
    \begin{equation}
        \Sigma( \phi^1(t,x) ) = \big\{ (t,x) \in M \mid \tau = \phi^1(t,x)  \big\}.
    \end{equation}
    Then we can rewrite \eqref{eq:avg-procedure-gasperini} in terms of the fields $S$ and $h$ as
    \begin{equation}\label{eq:avg-procedure-gasperini2}
        \langle S \rangle^\circ_\mathcal{D} (\tau) = \frac{ \int_{\mathcal{D}(\tau)} S(t_x, x) \sqrt{h (t_x, x) } d^3 x }{ \int_{\mathcal{D}(\tau)} \sqrt{h (t_x, x) } d^3 x },
    \end{equation}
    where $t_x$ is the time-coordinate value such that $\phi^1(t_x, x) = \tau$. We explicitly invoke subscript notation $t_x$ to indicate that when integrating over $\phi(t_x,x) = (\tau, x) \in \mathcal{D}(\tau)$ in \eqref{eq:avg-procedure-gasperini2}, the value of the time-coordinate $t_x$ varies and is thus \textit{not} constant in $D(\tau)$. Comparing $\langle \cdot \rangle^\circ$  written in the form above with $\langle S \rangle^{(3)}$, we conclude that they are fundamentally different as they average over different points in the spacetime: $(\tau, x) \neq (t_x, x)$ as generally $\phi^1(t_x, x) \neq t_x$. Figure \ref{fig:nonequivalent-averages-transformation} intuitively depicts the transformation $\phi$.

    Thus, when comparing $\langle S \rangle^\circ_{\mathcal{D}} (\tau)$ to $\langle S \rangle^{(3)}_{\mathcal{D}} (\tau)$, the integration region $\mathcal{D}(\tau)$ is the same, but the scalar fields $S$ and $\sqrt{h}$ under the integral are evaluated at different spacetime points. Hence, generally, for a scalar field $S:M \to \mathbb{R}$
    \begin{equation}
        \langle S \rangle^\circ_\mathcal{D} (\tau) \neq \langle S \rangle^{(3)}_\mathcal{D} (\tau).
    \end{equation}
    This, furthermore, proves that these two averaging procedures are not invariant under foliation-gauge transformations, and thus \eqref{eq:avg-procedure-gasperini} as well. 
    
   From the perspective of the integration domain, we do not weigh the field contributions in $\mathcal{D}$, but the transformed spacetime points ending up in $\mathcal{D}$ by the transformation $\phi$. This is counter-intuitive to what the foliation is in the first place: to fix the spacetime points we consider as `spatial' from which we induce the spatial region $\mathcal{D}$. This supports the mathematical reality that $\widehat{S}$ cannot be identified with $\mathcal{D}$, as the averaging construction implicitly induced the cosmological equivalence class such that $S$ is identified with $\mathcal{D}$.
   
    Formally, we pick beforehand a representation $(M, g, \{F_i\})$ from its equivalence class. As was proven above, there is a one-to-one correspondence between $M$ and the foliation $\Sigma$. Thus, scalar field $S \in \{F_i\}$ is identified with $\Sigma$. In the context of the foliation-gauge transformation $\phi$ considered above, the cosmology representation is mapped
    \begin{equation}
        \big( \Sigma, g, \{ F_i \} \big) \longmapsto \big( \phi(\Sigma), \phi_* g, \phi_* F_i \big).
    \end{equation}
    Specifically, $\phi_* S$ is identified with $\phi(\Sigma)$ and not with $S$ itself. Averaging $S$ over slices of the foliation $\phi(\Sigma)$ is thus unambiguous.

    To address the question whether a gauge-invariant averaging can be constructed, that is suitable for cosmic structure analysis, we turn to the theory of gauge-invariant perturbations.

\section{Gauge-invariant perturbative averaging}\label{sec:perturbative-avg}
Our treatment of foliation-gauges made the peculiarities of scalar averaging over cosmic structure apparent. Here, we provide a  solution to this problem in the context of standard cosmological simulations. To construct an appropriate averaging procedure, we build on the formalism of gauge-invariant perturbation theory, following \citet{bardeen1980gauge}.

By this perturbative approach, we can write any scalar field $S:M \to \mathbb{R}$ as a decomposition,
\begin{equation}
    S(t, x) = S^0(t) + \delta S^* (t, x),
\end{equation}
where perturbation $\delta S^*$ is invariant under scalar-preserving gauge-transformations. The decomposition is unique\footnote{Note that there is gauge freedom in setting conditions for constructing $\delta S^*$. However, this gauge freedom does not diminish the decomposition of $S$.} in its image for a specified background $S^0$. We define the \textit{perturbative averaging} of $S$ with background $S^0$ in domain $\mathcal{D} \subset \Sigma_t$ by
\begin{equation}\label{eq:perturbative-avg}
    \langle S \rangle^*_{\mathcal{D}} (t) := S^0(t) + \frac{1}{V} \int_{\mathcal{D}} \delta S^*_t(x) \sqrt{h(x)} \; d^3 x,
\end{equation} 
with $\delta S^*_t(x) := \delta S^* (t,x)$ and $V=\int_{\mathcal{D}} \sqrt{h(x)} d^3 x$ the volume of $\mathcal{D}$.

The invariance of \eqref{eq:perturbative-avg} under scalar-preserving gauge-transformations can be proven as follows. We assume the background $S^0$ to be invariant. Let $\varphi$ be such a gauge-transformation, whose effect on tensor fields is denoted by a tilde. The integral over the perturbation $\delta S^*_t$ transforms as
\begin{align}\label{eq:perturbative-avg-transformation}
    \int_{\mathcal{D}} \delta S^*_t (x) \sqrt{h(x)} \; d^3 x \longmapsto &\int_{\mathcal{D}} \widetilde{\delta S_t^*}(x) \sqrt{\widetilde{h}(x)} \; d^3 x \\
    = &\int_{\mathcal{D}} \delta S_t^* (x) \sqrt{h\big( \varphi^{-1}(x) \big)} \; \bigg| \frac{\partial \varphi}{\partial x}\bigg|^{-1} \; d^3 x,
\end{align}
using the fact that $\delta S_t^*$ is gauge-invariant. Making a local coordinate transformation $x \mapsto \bar{x} := \varphi^{-1}(x)$ gives
\begin{align}
     \int_{\mathcal{D}} \delta S^*_t (x) \sqrt{h(x)} \; d^3 x   \longmapsto &\int_{\mathcal{D}} \delta S^*_t \big( \varphi(\bar{x}) \big) \sqrt{h(\bar{x})} \; d^3 \bar{x} \\
     = &\int_{\mathcal{D}} \widehat{\delta S^*_t} (\bar{x}) \sqrt{h(\bar{x})} \; d^3 \bar{x}, \label{eq:pert-avg-last-transform}
\end{align}
where we denoted $\delta S^*_t \mapsto \widehat{\delta S^*_t}$ to be the change under $\varphi^{-1}$. Since the inverse $\varphi^{-1}$ is also a gauge-transformation, we see that \eqref{eq:pert-avg-last-transform} is unchanged. Hence, the perturbative averaging \eqref{eq:perturbative-avg} is invariant under gauge-transformation $\varphi$. A coordinate-independent argument as in Section \ref{sec:coord-indep-argument} is straightforwardly derived.

For cosmological modelling, this can be of particular interest. Using a spatial-independent background for scalar fields is standard practice in cosmology. Having a unique way to describing the deviations from the background, such as we described above, leads to a standard way of quantifying averaged deviations in local regions of the cosmic web. This makes the perturbative averaging \eqref{eq:perturbative-avg} in line with the concordance model of cosmology, and applicable to standard simulations.

\section{Conclusion}\label{sec:conclusion}
In the past few decades, extensive analyses of cosmic inhomogeneities and their dynamics influencing the behavior of large-scale structures have been undertaken. Most of these backreaction studies utilize either a perturbative approach or a spatially averaging method. We have demonstrated that backreaction exhibits a nested and multi-scale character, which refutes the prominent perturbative argument presented by \citet{green2011framework}. Their claim, asserting the insignificance of backreaction on large scales, has been shown to be untenable. One promising direction for further research is to adopt the perturbative formalism for studying backreaction effects on inhomogeneous structures across various clustering scales.

While scalar averaging approaches, which try to quantify backreaction, have shown numerous promising results in the literature, several mathematical challenges have been overlooked. We have provided a rigorous mathematical treatment on foliations and its connection to gauges in relativity. This analysis has brought to light that standard scalar averaging in a cosmological setting is intrinsically gauge-dependent. For instance, the ambiguity arises in the usage of 4-averaging with the 4-metric tensor and 3-averaging with the spatial metric. Although they are equivalent up to foliation-gauge changes, they are not invariant under these transformations.

To address the impact of averaging dependence, we have illustrated an example demonstrating that averaging can lead to artificial backreaction, particularly in the form of artificial acceleration terms. In response to this issue, we have discussed Gasperini's suggestion for a gauge-invariant averaging. However, we have clarified that the suggested averaging method is mathematically ill-defined, as the averaging constructions are derived from different representations of the cosmology.

To solve this issue, we adopted Bardeen's approach to construct a perturbative averaging method that is invariant under scalar-preserving gauge-transformations. Analyzing the restriction on the cosmological model of only considering gauge changes that preserve the scalar nature is a valuable next step for future research. 

Derivations and computations of backreaction effects with perturbative averaging in a gauge-invariant way will be presented in later papers.

\section*{Acknowlegdements}
We thank Robbert W. Scholtens for a reading of the manuscript and his useful comments.

\section*{ORCID iD}
Dave Verweg https://orcid.org/0000-0001-6163-2165 \\
Bernard J. T. Jones https://orcid.org/0000-0001-7700-3312 \\
Rien van de Weygaert https://orcid.org/0000-0001-8379-1263 \\

\bibliographystyle{mnras.bst}
\bibliography{ref}

\end{document}